\documentclass[authoryear,review]{elsarticle}

\usepackage[colorlinks=true]{hyperref}
\usepackage{graphicx}
\usepackage{amsmath}
\usepackage{diagbox}
\usepackage{natbib}
\usepackage{eurosym}
\journal{arXiv}

\begin{document}

\begin{frontmatter}


\title{Coevolution of theoretical and applied research: a case study of graphene research by temporal and geographic analysis}


%
%

\author[ntu_spms]{Ai Linh Nguyen\fnref{myfootnote}}

\author[ntu_spms]{Wenyuan Liu\corref{mycorrespondingauthor},\fnref{myfootnote}}
\cortext[mycorrespondingauthor]{Corresponding author}
\ead{wenyuan.liu@ntu.edu.sg}

\author[ntu_spms]{Siew Ann Cheong}

\address[ntu_spms]{Division of Physics and Applied Physics, School of Physical and Mathematical Sciences, Nanyang Technological University, 21 Nanyang Link, 637371, Singapore}

\fntext[myfootnote]{These authors contributed equally.}

\begin{abstract}
As a part of science of science (SciSci) research, the evolution of scientific disciplines has been attracting a great deal of attention recently.
This kind of discipline level analysis not only give insights of one particular field but also shed light on general principles of scientific enterprise.
In this paper we focus on graphene research, a fast growing field covers both theoretical and applied study.
Using co-clustering method, we split graphene literature into two groups and confirm that one group is about theoretical research (T) and another corresponds to applied research (A).
We analyze the proportion of T/A and found applied research becomes more and more popular after 2007.
Geographical analysis demonstrated that countries have different preference in terms of T/A and they reacted differently to research trend.
The interaction between two groups has been analyzed and shows that T extremely relies on T and A heavily relies on A, however the situation is very stable for T but changed markedly for A.
No geographic difference is found for the interaction dynamics.
Our results give a comprehensive picture of graphene research evolution and also provide a general framework which is able to analyze other disciplines.

\end{abstract}


\end{frontmatter}

\section{Introduction}

As an emerging field, science of science (SciSci) has attracted a great deal attention recently \citep{zeng_science_2017,fortunato_science_2018,wang_science_2021}.
In SciSci, science is treated as a complex system, which include ideas, papers, scientists, funding agencies and, more importantly, the connections among them.
Using methods from complex system and complex network, researchers have revealed many interest findings of Science from data.
They includes universal citation distribution \citep{radicchi_universality_2008}, scientists' career dynamics \citep{petersen_persistence_2012}, the role of team in science \citep{wuchty_increasing_2007}, just to name a few.

As a complex system, science ecosystem has obvious hierarchical structure: roughly speaking, science includes physical science and life science, physical science includes physics, astronomy, chemistry and earth science, physics includes mechanics, electromagnetism, thermodynamics, relativity, quantum physics and so on, electromagnetism itself also have its internal fine structure.
At each hierarchical level, scientific elements (people, ideas, papers) are closely connected within disciplines, and loosely connected between disciplines.
These naturally exist modular structure provide convenience to SciSci research since we can concentrate on particular discipline by assuming the influence from other disciplines is negligible.
This kind of discipline level analysis not only give insight of one particular field but also shed light on general principles of scientific enterprise.
Therefore, people's interest on discipline analysis has been growing rapidly.

One of the earliest extensive studies on this subject was done by  \citet{bettencourt_evolution_2011}.
The authors focus on the evolution of sustainability science, a new discipline emerged in 1980s.
By analyzing a large corpus of relevant publications, they found that sustainability science has been growing explosively since its advent in the 1980s.
This discipline has an unusual spatial distribution of its contributions: they are widely distributed in both developed counties and developing countries; the collaboration network has strong roots in national capital rather than traditionally more academic cities.
To capture the main themes that define the filed, the authors decomposed the corpus into traditional disciplines and found they are integrated management of human, social, and ecological systems from an engineering and policy perspective.
The work in \citet{bettencourt_evolution_2011} also revealed that the unification of sustainability science happened around the year 2000 by collaboration network analysis.

One recent discipline analysis look to artificial intelligence (AI) research \citep{frank_evolution_2019}.
The authors tried to figure out whether AI research and relevant social science fields keep pace with each other.
To answer this question, they used citation to track the communication between AI research and other fields.
They analyzed citation flows from 1950 to 2018 and found these flow are neither constant nor symmetric.
AI research cited philosophy, geography and art a lot in its early years, however current AI research cited mathematics and computer science most strongly.
On the other hand, other fields didn't cite AI research in proportion to its growing number of publications.
There is an attention gap between AI research and social science.

Here, we conduct a discipline analysis of graphene research, a relatively young field focus on single layer of carbon atoms arranged in a two-dimensional honeycomb lattice.
Since \citet{novoselov_electric_2004} seminal work, which helps Andre Geim and Konstantin Novoselov won Nobel Prize in Physics in 2010, the interest in graphene has grown explosively and even led European Commission to fund Graphene Flagship with \euro{}1 billion budget \citep{noauthor_graphene_nodate} and the creation of National Graphene Institute in UK \citep{noauthor_national_nodate}.
Therefore, it is important and beneficial to understand graphene research as a emerging research discipline.
It will not only increase our knowledge about scientific enterprise at discipline level, but also provide useful information of the whole discipline to PhD students, graphene scientists, universities and funding agencies.  

Generally speaking, there are two reasons that attract people to do graphene research: 1) graphene is very strange from a theoretical perspective. 
It is the first example of two-dimensional atomic crystal and it was thought that such structure is thermodynamically unstable.
The graphene's electronic properties are peculiar: electrons behave as massless relativistic particles.
2) graphene is very attractive from an applied perspective. 
It has many properties which are better than those in all other materials.
Hence, graphene has enormous potential to improve many applications, range from electronics to medicine.
And these two reasons guide people go to different directions and form two branches: theoretical branch and applied branch.
Although this division is well known in graphene research community, there is a lack of quantitative analysis from this perspective \citep{barth_graphene_2008,lv_bibliometric_2011,nguyen_golden_2020}.
To fill this gap, we collect graphene research literature and conduct a systematic analysis.
First we identify these two branches by their word frequency, then validate the division result with keywords.
We find that theoretical branch includes around one third of all papers and applied branch includes remaining two third.
We check how these two branches developed temporally and geographically.
The results show that yearly proportion of applied branches decreased from 2004 to 2007 and kept increasing since then.
The research contribution mainly come from a few regions, and geographic distributions are quite different in theoretical and applied research.    
Finally we study the interaction between two branches and find theoretical research extremely relies on past theoretical research while applied research heavily relies on past applied research.
Temporal analysis suggests that this self-dependence was very stable for theoretical research while grown significantly for applied research.
And this trend holds for all main regions of graphene research.

This paper is organized as follows.
In \autoref{sec:method} we describe our dataset and the framework to quantify and analyze the inside coevolution.
In \autoref{sec:result} we present our main results, namely, the internal structure of graphene research and linguistic validation, spatial and temporal analysis of graphene publications and their interdependence, and also discuss the limitations and implications.
\autoref{sec:conclusion} summarize our findings and suggests future perspectives.

\section{Methodology}\label{sec:method}

\subsection{Dataset}

To build the dataset, we chose ‘graphene’ as the topic keyword to search in the Web of Science Core Collection and obtained bibliographical records of 135,617 graphene-related journal papers in August 2018. 
These records have been used in our another paper \citep{nguyen_golden_2020} and interested readers can find more information about these records there. 
Web of Science may not cover every journal publication in this topic, however, given the wide coverage of Science Citation Index, most mainstream graphene papers should be included in our dataset. 

There are various document types in 135,617 records: articles, proceeding papers, reviews, meeting abstracts, etc. 
Since the primary focus of this paper is coevolution between theoretical and applied branches in graphene research, we only included \textbf{research articles} with DOI names and publication years for our analysis. 
There are 115,988 remaining records and all analyses in this paper are done with them if not mentioned otherwise.

\subsection{Co-clustering}\label{sec:co-clust}

To group these papers into clusters by their research topics, we applied a block diagonal co-clustering algorithm introduced by \citet{ailem_co-clustering_2015,ailem_graph_2016}, namely CoClus, to divided papers into a number of non-overlapping clusters with their characteristic words. 
It is well known that papers with different research topics tend to have different word frequency features and researchers have demonstrated that Coclus algorithm can effectively co-cluster document-word matrices \citep{ailem_graph_2016}. 
In this paper, we assume (a) the linguistic content inside the title and abstract is sufficient to tell the topic in each article, and (b) words that appear less than 0.01\% of all records have insignificant impact on the clustering process. 
Generally speaking, CoClus algorithm aims to partition the object set \textbf{I} of size P and corresponding attribute set \textbf{J} of size N into $g$ non-overlapping clusters with high in-cluster density and low cross-cluster density.
In our case, the goal is to split our data collection (papers and associated words) into $g$ groups and each group has a set of papers and a set of words.
These words are used more frequently in the papers belong to the same group and less frequently in the papers belong to other groups.

Firstly, every article is represented by a N-dimensions vector $a_i$:
\begin{equation}	
	a_i = (a_{w_1}, a_{w_2},...,a_{w_N}),
\end{equation}
where $N$ is the number of feature words we considered in this study and $a_{w_j}$ is the count of word $w_j$ in article $a_i$.
By concatenating all paper vectors together, we constructed a paper-word matrix A:
\begin{equation}
A = 
\begin{pmatrix}
a_1 \\
\dots \\
a_P
\end{pmatrix}
=
\begin{pmatrix}
a^1_{w_1} & \dots & a^1_{w_N} \\
\vdots & \ddots & \vdots \\
a^P_{w_1} & \dots & a^P_{w_N}
\end{pmatrix},
\end{equation}
where $a^P_{w_N}$ represents the total number of word $w_N$ in the title and abstract of article $P$.
Then, the algorithm tries to cluster matrices effectively by introducing a block seriation $C$:
\begin{equation}
	C = \{c_{pn}\}_{\substack{p=1,\dots,P\\n=1,\dots,N}},
\end{equation} 
where $c_{pn} = 1$ if object $p$ and attribute $n$ belong to the same cluster and $c_{pn} = 0$ otherwise.
\citet{ailem_co-clustering_2015} introduced a reformulated modularity as:
\begin{equation}
	Q(A,C) = \frac{1}{\sum_{p,n}a^p_n}\sum_{p=1}^{P}\sum_{n=1}^{N}\Big(a^p_n - \frac{\sum_{n}a^p_n\sum_{p}a^p_n}{\sum_{p,n}a^p_n}\Big)c_{pn}.
\end{equation}
And it turns out that partition with high in-cluster density and low cross-cluster density is equivalent to high $Q$ and the $C$ that returns largest $Q$ is the best partition we are looking for. 
The reformulated modularity has a linear dependence on $C$, therefore the co-clustering task can now be regarded as an integer linear programming problem. 
A python package CoClust \citep{role_coclust_2019} was used to find the partition $C$ that returns the largest reformulated modularity $Q$. 
Each paper or word has a cluster label $g$ in $C$ and we can easily construct any cluster by grouping papers or words with the same label $g$.
This is the main idea of CoClus algorithm and we refer the interested readers to \citep{ailem_co-clustering_2015,ailem_graph_2016}.

\subsection{Keywords by Z-score}

Co-clustering method can help us divide papers and words into several groups.
However, it is not a panacea for all research problems.
Apart from the technical perspective, the result of co-clustering heavily depends on how you abstract your research problem into mathematical form.
If the abstraction is not reasonable, then co-clustering may not return meaningful results.
Therefore it is very important to validate co-clustering result from different perspectives.
In this study, we built a keyword list for each cluster.
If the co-clustering works well, then one keyword list should have more ``theoretical'' words and another should have more ``applied'' words and this can be easily checked by anyone with basic science training.
Beyond the validation, these keywords also give us a comprehensible overview of graphene research.

There are many ways to extract keywords from corpus \citep{berry_text_2010}.
The most straightforward option for us is to count all words in each cluster and pick most frequent words as keywords. 
However this naive method may take general words like ``research'', ``study'', ``found'', ``result'' or even ``graphene'' which are extensively used in most papers.
Such words are not informative as keywords for graphene research topics.
Therefore, we use statistical significance of word occurrence to measure the correlation between words and clusters. 
The words with strong correlation are the keywords for that cluster.

To measure the correlation between word $w_j$ and cluster $I_k$ (with $N_k$ papers), 
we first count the number of papers in $I_k$ that contain $w_j$ in their titles and abstracts, refer to it as $\mu_{w_j,k}$. 
We also measure frequency of $w_j$ in the whole data collection and denote it as $P_{w_j}$.
By assuming the distribution of $\mu_{w_j,k}$ follows a binomial function, the expected mean and variance of the number of records in $I_k$ that contain $w_j$ are:
\begin{gather}
\overline{\mu}_{w_j, k} = N_k \cdot P_{w_j}, \\
\sigma_{w_j, k} = N_k \cdot P_{w_j} \cdot (1 - P_{w_j}).
\end{gather}
Then, the z-score for word $w_j$ can be defined as:
\begin{equation}
	z_{w_j} = \frac{\mu_{w_j, k}-\overline{\mu}_{w_j, k}}{\sigma_{w_j, k}}.
\end{equation}
Apparently, a word that appears more often in a cluster rather than any others will have a high z-score in that cluster. 
On the other hand, words with high z-score do not necessarily appear often, as long as its frequency is higher than expected value.
The top z-score words should be able to illustrate the topic of each community and we use them as keywords for each cluster.

\subsection{Regional credit}\label{sec:credit}

As a high competitive field, many regions keep a important position for graphene when they make their research programme.
For example, European Commission fund Graphene Flagship with \euro{}1 billion budget \citep{noauthor_graphene_nodate}, UK government fund National Graphene Institute with \pounds38 million \citep{noauthor_national_nodate} and South Korea government announced the ``Technology Roadmap for Promoting Commercialization of graphene (2015–2020)'' in 2015 \citep{}.
Since regions have different research traditions, resource and targets, it is not surprise that they may has different aims and preferences in graphene research.
To examine this hypothesis, we counted authors' affiliations of each paper and calculated regional credit accordingly.
More specifically, if all authors of paper $P$ only have affiliations in region $R$, then $R$ has full credit for this paper $P$, that is 1.
On the other hand, if paper $P$ is result of international collaboration, we first split the credit among all authors evenly.
Then, for each author, we split his/her credit among all his/her affiliations uniformly.
Finally the credits associated with each affiliation are added together to get regional credits to that paper. 
They can be fractions for regions.
For instance, paper $P$ has three authors $A_1$, $A_2$ and $A_3$.
$A_1$ has two affiliations, one in country $C_1$ and the other in country $C_2$.
$A_2$ has only one affiliation in country $C_3$.
$A_3$ has two affiliations both in $C_2$.
Then $C_1$ has credit $\frac{1}{6}$, $C_2$ has credit $\frac{1}{2} = \frac{1}{6} + \frac{1}{3}$ and $C_3$ has credit $\frac{1}{3}$.

\subsection{T/A dependency}

After graphene research literature be divided into theoretical and applied branches, a question emerges naturally: how did theoretical branch and applied branch influence each other and shape current graphene research? 
Although the existence of interplay between them is apparent, the extend and strength of interaction are far from obvious. 
Generally, theoretical research provides guideline for applied research and applied research contributes proving ground for theoretical research . 
Meanwhile, theoretical research inspires followed discussion of theoretical questions and applied research stimulate subsequent applied study for the common interest.
Given those complicated interactions, a quantitative method is needed to measure the influence between theoretical research and applied research.
Only then we can answer that question from our dataset.

In this study, we use citation to capture influence: paper A cites only theoretical papers means only theoretical research influence A, paper B cites only applied papers means only applied research influence B, paper C cites theoretical and applied papers means both theoretical and applied research influence C. 
At first glance, it is tempting to simply use proportion of reference to measure influence: if paper D cites 3 theoretical papers and 7 applied papers, then theoretical research has 30\% influence and applied research has 70\% influence over paper D. 
However, this method oversimplifies the process of knowledge accumulation: the 3 theoretical papers in D’s references may cite some applied papers and the 7 applied papers in D’s references may heavily depend on early theoretical works. 
Just counting references will miss this information, therefore, can not accurately reflect the interaction between T and A.

Inspired by persistent influence in \citet{della_briotta_parolo_tracking_2020}, we introduce a dependency factor pair $(D_T^n, D_A^n)$ of paper $n$ to describe its dependency on theoretical research ($D_T^n$) and applied research ($D_A^n$). 
A paper with a pair $(0.7, 0.3)$ means it receives 70\% influence from theoretical research and 30\% influence from applied research.
$(D_T^n, D_A^n)$ satisfies following conditions: $0 \le D_T \le 1$, $0 \le D_A \le 1$, and $D_T + D_A = 1$ for obvious reason.
To avoid the oversimplification mentioned in previous paragraph, we split whole dependency into two parts: direct dependency ($d_{T/A}$) and indirect dependency ($i_{T/A}$).
The direct dependency describes paper's reference proportion of T and A, while the indirect dependency capture those references' dependency.
The direct dependency is very straightforward: for any paper, we only need to check its reference: if $x\%$ of them are theoretical papers, then $d_T = x\%, d_A = 1-x\%$.
We use indirect dependency to reflect explicit influence: paper A may only cites applied papers, however those applied papers cite many theoretical papers, therefore paper A benefits from theoretical research indirectly and this influence is captured by $i_{T/A}$.
For paper $n$, its indirect dependency is defined as the average of all its references' dependency factor, that is to say,
\begin{subequations}\label{eq:indirect_dependency}
\begin{gather}
	i_T^n = \frac{1}{\|refs\ of\ n\|}\sum_{m \in refs\ of\ n} D_T^m, \\
	i_A^n = \frac{1}{\|refs\ of\ n\|}\sum_{m \in refs\ of\ n} D_A^m.
\end{gather}
\end{subequations}
By combining the direct and indirect dependency, our definition can reflect paper's reliance on theoretical and applied research more accurately and comprehensively.

\begin{figure}
	\includegraphics[width=\columnwidth]{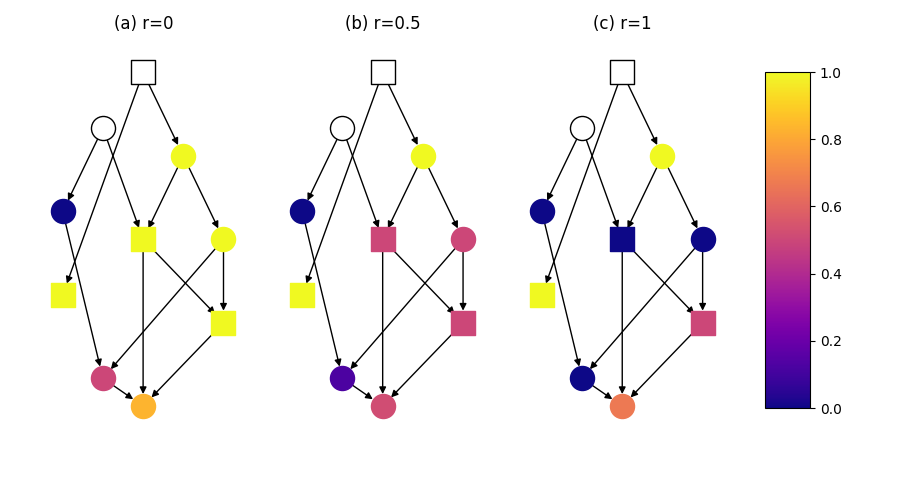}
	\caption{Illustration of dependency factor in citation network. In all panels, we plot the same citation network where nodes are papers and each edge represents a citation. The arrow corresponds the flow of influence, i.e., from a cited paper to a citing paper. The square nodes are theoretical papers and the circle nodes are applied papers. All nodes are colored according to their $D_T$ values, which are calculated using \autoref{eq: dependency}. Root papers are left blank as their $D_T$ are undefined. The control parameter $r$ is set as 0 in (a), 0.5 in (b) and 1 in (c).}\label{fig:depedency}
\end{figure}

After obtaining $(d_T^n,\ d_A^n)$ and $(i_T^n,i_A^n)$, the direct and indirect dependency of a paper $n$ on T and A respectively, the overall dependency on T and A, $D_T^n$ and $D_A^n$ can be expressed in the following way:
\begin{subequations}\label{eq: dependency}
\begin{gather}
	D_T^n = r \cdot d_T^n + (1 - r) \cdot i_T^n, \\
	D_A^n = r \cdot d_A^n + (1 - r) \cdot i_A^n,
\end{gather}
\end{subequations}
where $r$ is the control parameter determining the mix ratio of direct and indirect dependency with constrain $0 \le r \le 1$.
By setting $r = 1$, dependency factor will reduce to only direct dependency, that is to say dependency factor will only reflect paper's reference proportion in T/A.
On the other hand, dependency factor will reduce to only indirect dependency with $r = 0$ and reference proportion in T/A will not have explicit effect.
These extreme cases are the oversimplification problem we have discussed in previous paragraphs.
Through introducing $r$, we are able to get over these oversimplification without loss of generality.
The effect of $r$ is shown in \autoref{fig:depedency}.
We discussed the numerical effect of $r$ is in SI and its value is set to $0.5$ in this study if not mentioned otherwise.
It is worth noting that if a paper does not cite any paper in our dataset (only be cited), we can not define its dependency factor.
We call such papers as "root papers" and discuss about them in SI.

\section{Results and discussion}\label{sec:result}

We start our study by identifying theoretical and applied papers in our data collection. 
Given the complexity of research, this dichotomy may miss a little information since some papers are hard to be categorized. 
However, this classification is well accepted by graphene research community and our results are in good agreement with this convention. 
Dividing graphene literature into two branches can capture the most significant heterogeneity inside graphene research. 
Therefore, we use theoretical/applied dichotomy through this paper and call them T and A for short. 

It is well known that word frequency changes significantly from field to field. 
We assume theoretical graphene papers tend to use a set of words frequently while applied graphene papers tend to use another set of words frequently and these two sets of words are distinct. 
Based on this assumption, we first counted all words in titles and abstracts of 115,988 papers through standard natural language processing techniques (tokenization, stopword filtering, stemming and so on). 
The words with frequency larger than 0.01\% are selected as feature words and we have 12328 of them. 
Every article is represented by a 12328-dimensions vector $(a_{w_1},a_{w_2},\dots,a_{w_{12328}})$ and ${a_w}_j$ is the count of word $w_j$ in that article. 
Combining all paper vectors together, we constructed a $115988\times12328$ paper-word matrix. 
Then the CoClus method in \autoref{sec:co-clust} was applied to extract no-overlapping clusters. 
In CoClus we must specified number of cluster $k$ first to find best partition with this $k$.
Normally, the number of clusters is unknown at the beginning. 
The common protocol is to repeat this process with different $k$ and choosing $k$ with highest modularity as the result.
We run CoClus algorithm with $2 \le k \le 9$ and found the modularity peaks at two values $k=4$ and $k=6$, see supplementary information for more details.

However, the cluster structure in paper-word network is quite fuzzy.
More effort is needed to get the reasonable partition.
By comparing best partitions under different $k$, we noticed they share a common feature: one group repeatedly occurred in most partitions while other groups are not stable.
It suggests that there exist a distinct boundary between that group and others while other detected structures are more or less the ``overfitting''. 
To show the stability of this ''hyperstructure'', we compare the partitions with two highest modularity, namely, $k=4$ and $k=6$.
In case of $k=4$, We named the stable group as I and other groups as II, III, IV for convenience.
For the same reason, we named the stable group as 1 and other groups as 2, 3, 4, 5, 6 in case of $k=6$.
Since the sum of group II, III and IV is the complement of group I, we call it group I' and also call the sum of group 2, 3, 4, 5, 6 as group 1'.
We find group I has 43323 papers, group 1 has 42798 papers and they have 40406 papers in common; group I' has 72665 papers, group 1' has 73190 papers and they have 70293 in common.
The situations are very similar for other $k$ values.
Therefore, to avoid the ``overfitting'', we use the ``hyperstructure'' with $k=4$, namely group I and I' as the partition result. 

\begin{figure}
	\includegraphics[width=\columnwidth]{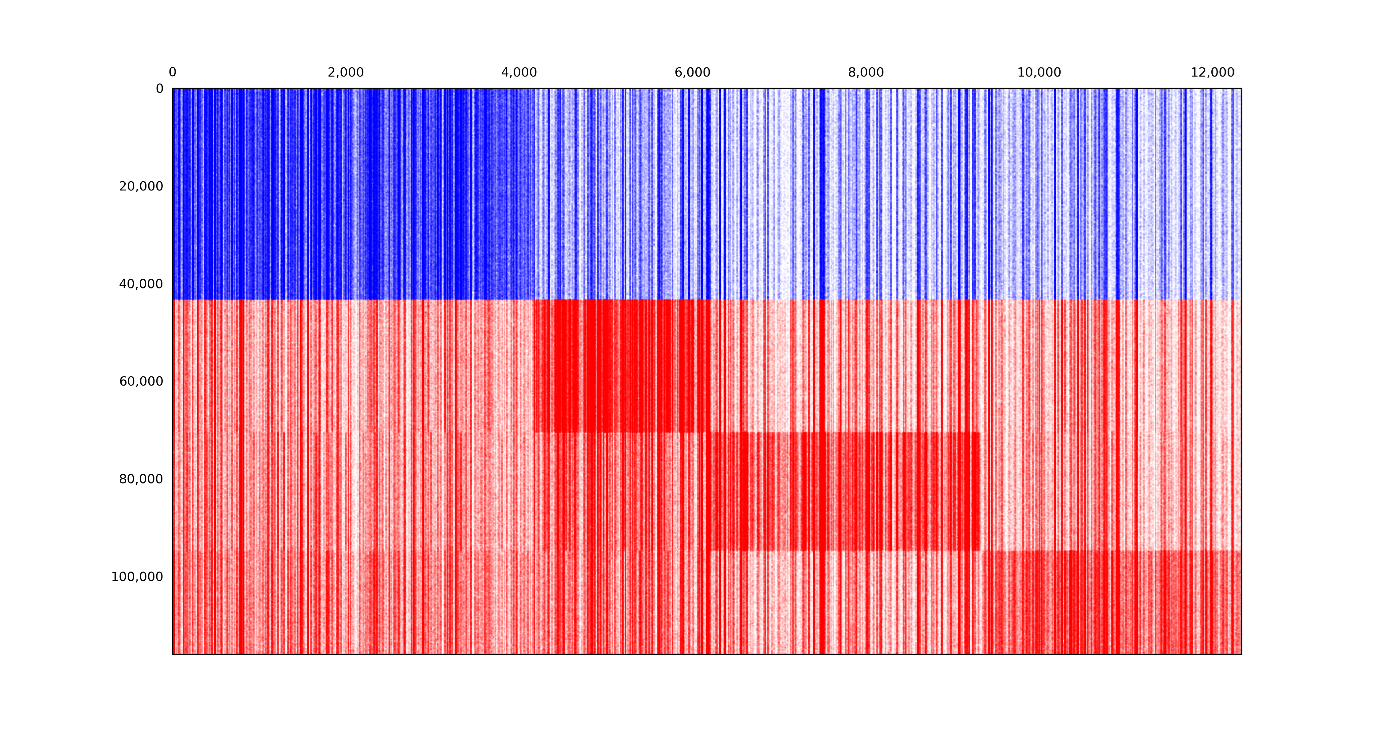}
	\caption{Heatmap for word distribution in our data collection. There are 115988 rows and 12328 columns. Each row represents a paper and each column represent a word. A filled block means that word appears in corresponding paper, otherwise we leave it empty. For the purpose of comparison, theoretical papers are colored blue and applied papers are colored red.}\label{fig:co-clust}
\end{figure}

Using the keyword extraction method, we found the stable group is about theoretical research and we call it T.
Other three groups under $k=4$ are (1) synthesis and functionalization, (2) supercapacitor, and (3) sensor.
That also explain the fuzzy boundary between them since they are closer to each other than they are from theoretical research.
Since they are application-oriented research comparing with T, we merged them into one cluster as A.
There are 72665 papers, 8162 words in group A and 43323 papers, 4166 words in group T.
We visualized the partition in \autoref{fig:co-clust}: paper vectors in T are colored blue and paper vectors in A are colored red.
We also sorted columns to make first 4166 columns represent words in group T and remaining 8162 columns represent words in group A.
As illustrated by \autoref{fig:co-clust}, it is apparent that first 4166 words are used more frequently in group T and other words are used more frequently in group A since these areas are darker than adjacent blocks.
And this is exactly the aim of co-clustering: high in-cluster density and low cross-cluster density.
It is worth noting that there are three subclusters inside A (these blocks are darker than other red area).
They are more specific topics (1) synthesis and functionalization, (2) supercapacitor, and (3) sensor, respectively.

\begin{figure}
	\includegraphics[width=\columnwidth]{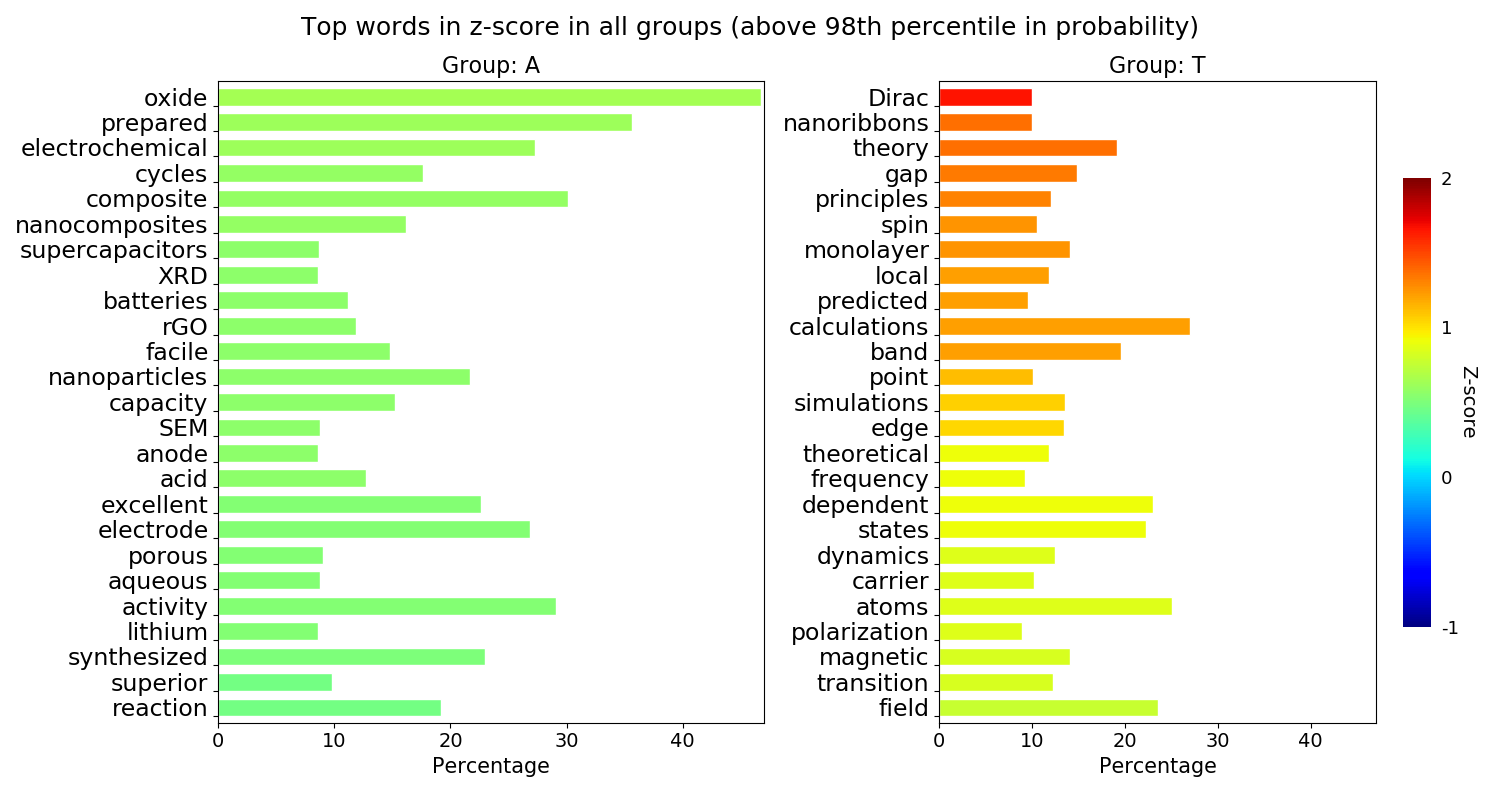}
	\caption{Top keywords by z-score in group A and T. Each bar’s length is proportional to the word frequency in that group and color is based on z-score.}\label{fig:keywords}
\end{figure}

To validate the clustering result and have an intuitive picture of research topic, we built a keyword list for each cluster.
To avoid statistical fluctuations, we only considered the words those are in top 2\% sorted by frequency.
The results are not sensitive to the quantile we chosen here as long as we dropped rare words (large fluctuations).
These keywords are ranked by their z-scores and top 25 keywords for T and A are shown in \autoref{fig:keywords}.
The lists are very informative: list A covers hot terms in applied graphene research, like electrochemical, cycles, supercapacitors, batteries, electrode, lithium and so on; list T covers key concepts in theoretical graphene research, such as Dirac, gap, spin, calculations, band, point, states and so on.
Therefore, we can conclude confidentially that co-clustering method successfully divide graphene research literature into theoretical branch and applied branch.
We also give two more extensive word clouds for T and A in SI. 

\begin{figure}
	\includegraphics[width=\columnwidth]{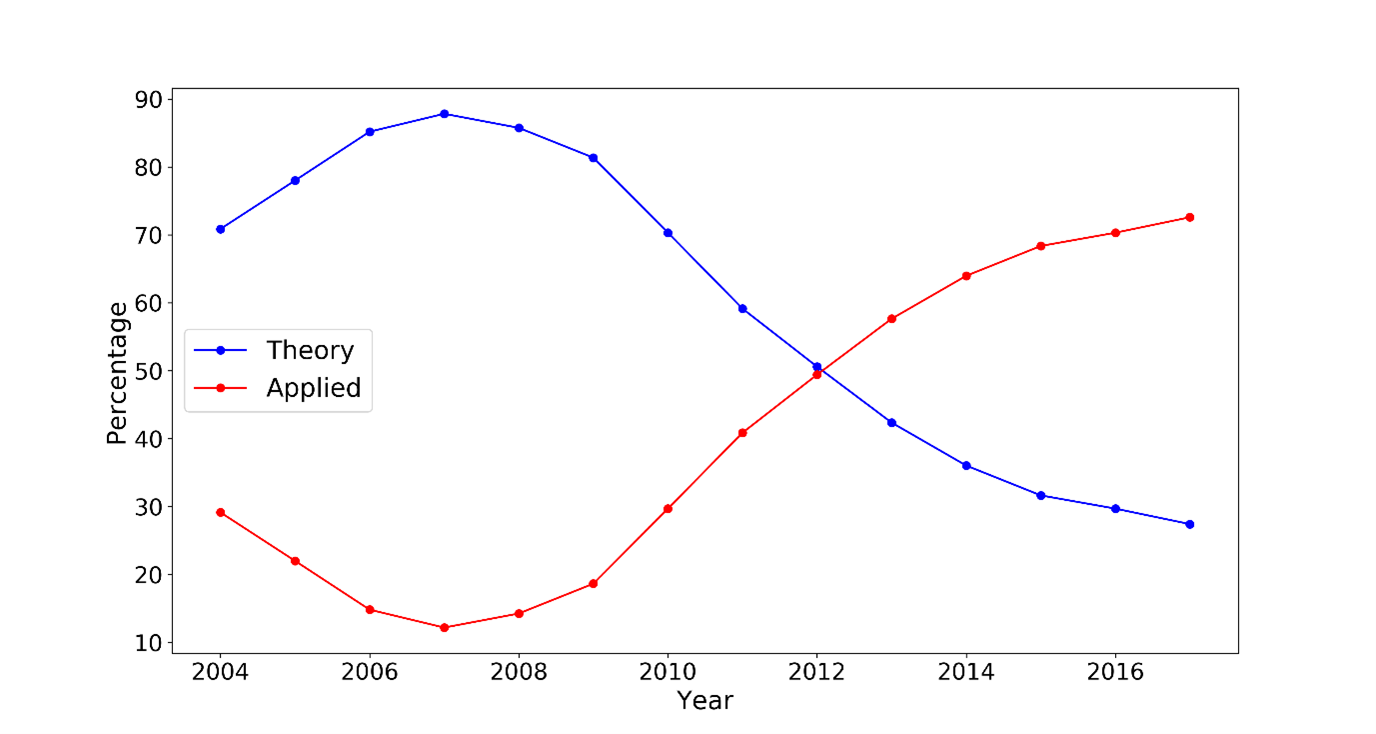}
	\caption{The yearly percentage of T and A from 2004 to 2017.}\label{fig:time_trend}
\end{figure}

Based on the co-clustering result, we first analyze the proportion of T and A.
Among all graphene research papers, 62.6\% belong to group A and 37.4\% belong to group T.
It suggests that graphene research attract attention from both theoretical and applied perspectives and both have made indispensable contribution to this field.
Furthermore, this ratio is not constant over time.
As shown in \autoref{fig:time_trend}, the proportion of T increased from 70\% to around 90\% during 2004-2007, then gradually decreased to lower than 30\% in 2017.
(Our temporal analysis focus on papers since 2004 because that is the year graphene got global attention and papers about graphene before 2004 are rare, only about 0.5\% in our dataset.)
These curves indicate that at the early stage of graphene research, theoretical branch played a dominant role and gained even more popularity until 2007.
After that applied branch grown relatively faster and became the majority after 2012 and this trend kept until 2017.
The reason behinds this process is not clear.
Our speculation is that after \citet{novoselov_electric_2004} seminal paper, graphene research attract a great deal of attention.
At that time, this area was still in it's infancy and many theoretical questions remained to be answered, while preparation of graphene was difficult and expensive, applied research is only limited to few labs.
So researchers published more theoretical papers than applied papers.
As time elapsed, people gained more understanding of graphene, low-hanging fruit has been picked and theoretical questions became more difficult and time-consuming.
On the contrary, technical advancement make preparation of graphene easier. 
It lowered research barrier and allowed more scientists joined the applied research.
Also, more understanding of graphene inspires more application scenarios, which motivate more applied research.
All of these factors together shape the curves in \autoref{fig:time_trend}. 
Although this hypothesis remains to be validated, our finding that proportion of group A has increased steadily since 2007 provide a big picture of graphene research ecosystem for researchers, companies and funding agencies.

\begin{figure}
	\includegraphics[width=\columnwidth]{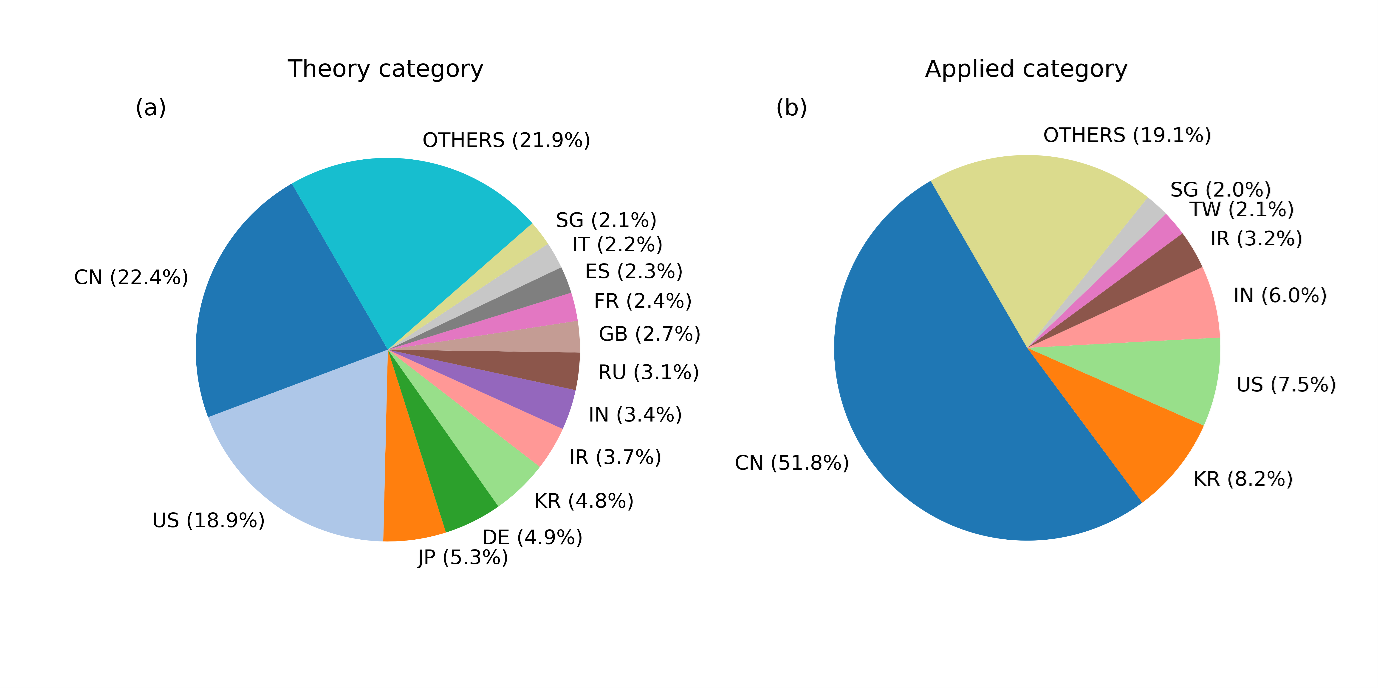}
	\caption{The top regions of contribution in T and A. Only regions with at least 2\% are shown. See Appendix A for region codes.}\label{fig:region_contribution}
\end{figure}

Beside temporal evolution, we also studied geographic distribution of graphene research.
As a highly actively field with enormous economic potential, it attracts scientists and engineers all over the world. 
Affected by tradition, manpower and funding policy, regions may have different aims and preferences in graphene research.
To validate this conjecture, we calculated regional credit using method in \autoref{sec:credit}.
By summing all papers' credit distribution, we are able to measure each region's contribution in the whole field.
As shown in \autoref{fig:region_contribution}, Mainland China is the topmost player in both theoretical and applied research, with 22.4\% share in T and 51.8\% share in A.
That means Mainland China's contribution in applied research is even more than the sum of all other regions.
The United States is also a big player with 18.9\% share in T and 7.5\% share in A. 
In contrast to Mainland China, US has more share in T than A.
It suggests that Mainland China is more focus on applied graphene research while The United States takes a more balanced position.
Furthermore, the composition of \autoref{fig:region_contribution} (a) and (b) reveal a subtle difference between theoretical and applied research: there are 13 regions with at least 2\% contribution in T while only 7 such regions in A.
It suggests that applied research is less geographically diverse than theoretical research.
The reason behinds is complicated, may due to economic factor, funding policy, research culture and so on.
We leave it for future research.

\begin{figure}
	\includegraphics[width=\columnwidth]{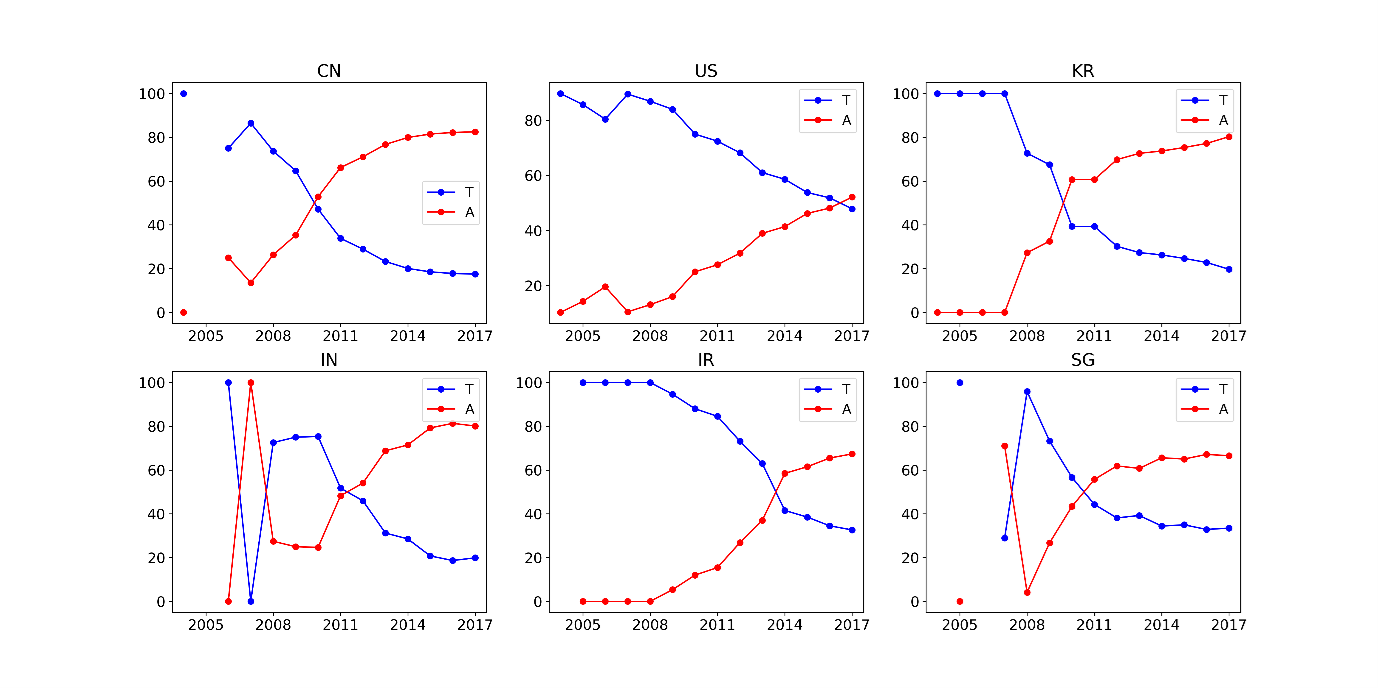}
	\caption{The yearly proportion of T and A in Mainland China, The United States, South Korea, India, Iran and Singapore from 2004 to 2017. Curves break when that region has not paper that year.}\label{fig: region_trend}
\end{figure}

So far we have studied the graphene research temporarily and geographically.
If this two dimensions are combined together, we are able to analyze evolution of graphene research in particular regions.
More concretely, for a given region in a given year we can calculate its credit in T  ($C_T$) and A ($C_A$) by method in \autoref{sec:credit} using publications only in that year.
Then the yearly proportion of T and A of that region in that year can be calculated as $\frac{C_T}{C_T + C_A}$ and $\frac{C_A}{C_T + C_A}$.
And we plot yearly proportion of T/A for six regions in \autoref{fig: region_trend}.
These six regions have at least 2\% share in both theoretical and applied research (see \autoref{fig:region_contribution}), therefore, are considered as top players in graphene research.
The curves in \autoref{fig: region_trend} can be thought of as regional version or components of \autoref{fig:time_trend} and it shows that all six regions follow the  same general trend in \autoref{fig:time_trend}: gradually decrease of theoretical research and increase of applied research.
However, that shift occurred slower in The United States than in other regions.
It suggests that research community in The United States still put considerable resource in theoretical research.
This finding is very important to understand graphene research competition among regions and make funding policy. 
    
Like most, if not all, science research fields, both theoretical and applied branches are indispensable parts of graphene research.
Each of them has its own mission and focus.
At the same time, they both rely on the communication with each other: theoretical research gets feedback from applied research, applied research receives guideline from theoretical research.
This inside coevolution is extremely important for any science research fields.
If this coevolution mechanism does not work well, that research field will experience certain difficulty to move forward, like Aristotelian Physics or science in ancient non-western civilizations.
Given such importance, we quantified the inside coevolution in graphene research in terms of interplay between T and A.
The dependency factors were computed for all papers, then the average values were calculated for papers in group T and A respectively.
As shown in \autoref{tab: dependency}, both T and A rely more on itself than others.
However, the difference is obvious: T is critically depend on T (90\%) while A is relatively rely on A (69\%).
This result illustrates a remarkable difference between T and A: theoretical research is mostly driven by other theoretical research, on the other hand applied research pays fairly high attention to theoretical research.
On possible interpretation of this difference is that: theoretical graphene research have achieved significant progress and its current effort is beyond existing technology, while applied research benefits a lot from theoretical research and as the result A cites T, directly and indirectly, a lot.
More evidences are need to verify this explanation, however, given the fact that in history of science theory is ahead of application at most of time , we believe it is a plausible explanation.

\begin{table}
\centering
\caption{The average dependency of T and A groups on T and A respectively.}\label{tab: dependency}
\begin{tabular}{|c|c|c|}
\hline
\diagbox{Group}{Dependency} & Theoretical (T) & Applied (A) \\
\hline
Theoretical (T) & 0.90 & 0.31 \\
\hline
Applied (A) & 0.10 & 0.69 \\
\hline
\end{tabular}
\end{table}

\begin{figure}
	\includegraphics[width=\columnwidth]{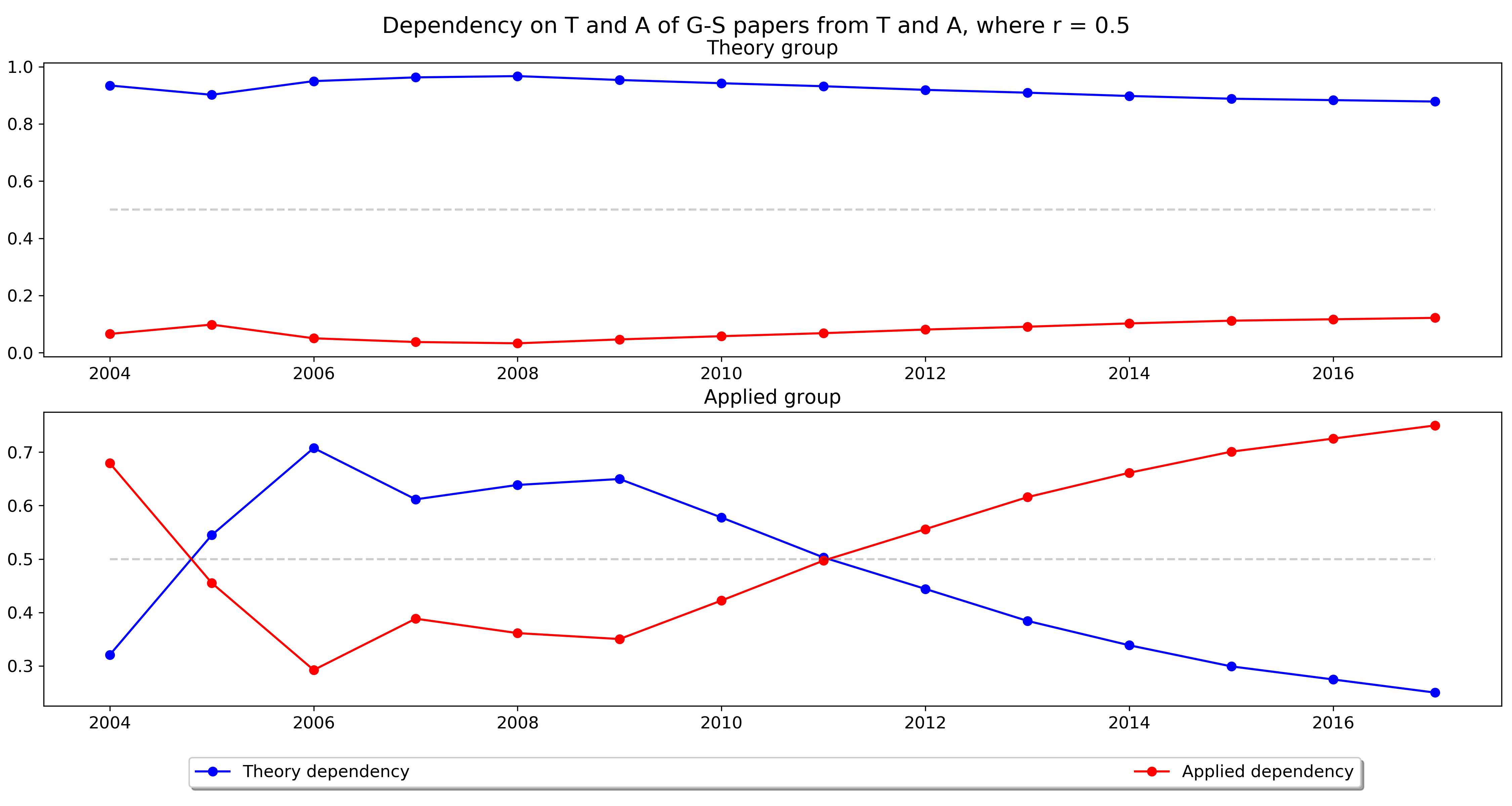}
	\caption{The yearly average dependency of T (top panel) and A (down panel) on T (blue) and A (red) respectively.}\label{fig: time_dependency}
\end{figure}

The numbers in \autoref{tab: dependency} is aggregated results of papers published in all years.
Even though they provides many insights for us, the temporal information is lost. 
To get a better understanding of coevolution process, we calculated the average values of T/A dependency for T/A papers in each year and plot them in \autoref{fig: time_dependency}.
As shown in that figure, dependency curves behavior significantly different between theoretical and applied group: the curves in T is very stable, while the curves in A underwent dramatic changes during that time.
From 2004 to 2006, A's dependency on T increased from 0.3 to 0.7 and gradually decreased afterward. 
That is to say applied graphene research is mainly driven by theoretical research at early stage and become more motivated by it self as time goes on.
This result suggests that the inside coevolution is not a ``static equilibrium'', but rather a ``dynamic process''.
Not only proportions of T and A change with time  (\autoref{fig:time_trend}), the interaction between them also changes.
More work is needed to fully understand the reason behind these changes, and our finding can serve as the basis for further research. 

\begin{figure}
	\includegraphics[width=\columnwidth]{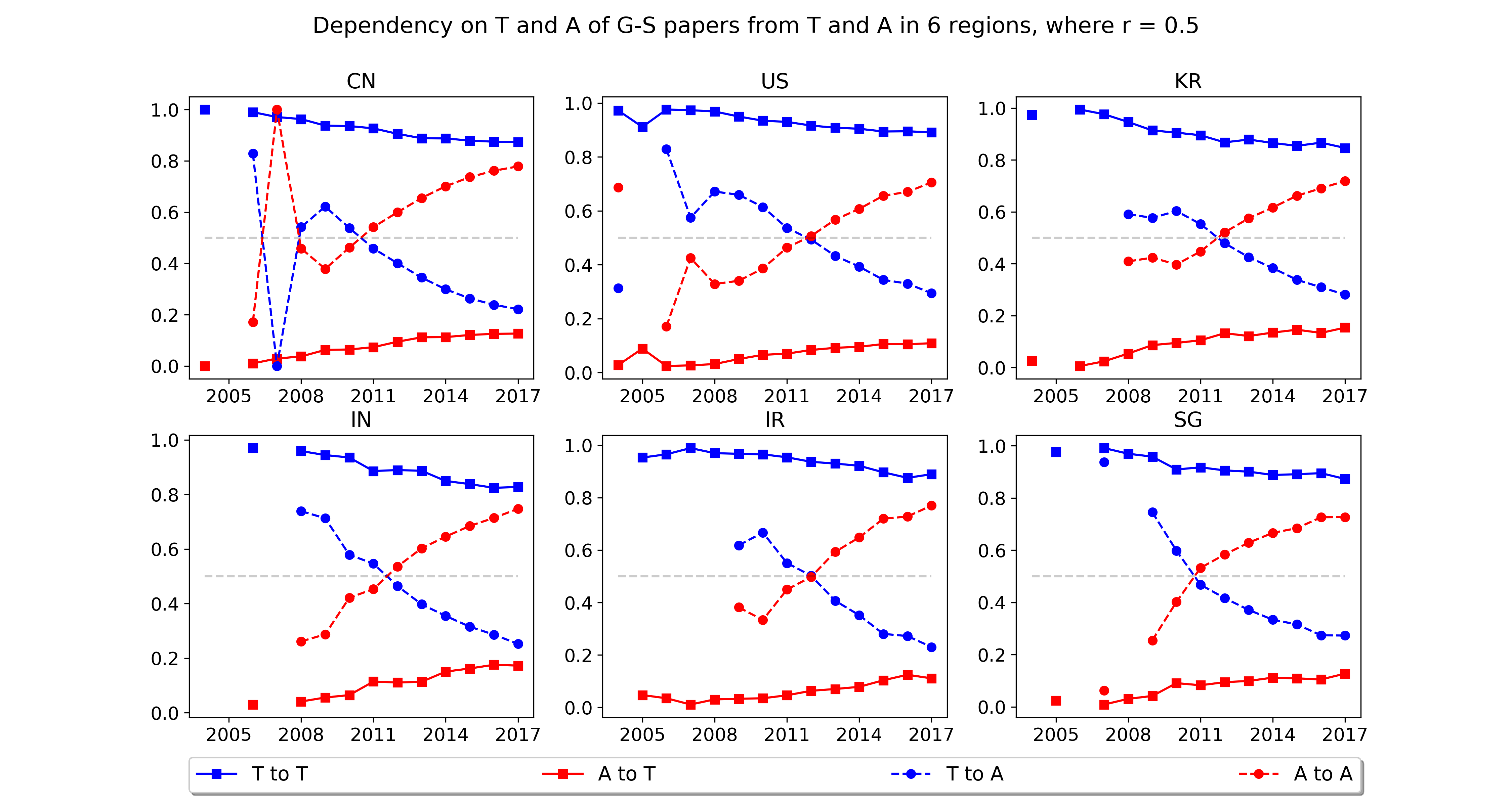}
	\caption{The yearly average dependency of T (square) and A (circle) on T (blue) and A (red) in Mainland China, The United States, South Korea, India, Iran and Singapore from 2004 to 2017. Curves break when that region has not T/A paper that year.}\label{fig: region_dependency}
\end{figure}

We have already shown the geographically difference in \autoref{fig: region_trend}.
Does such effect also happen in T/A dependency?
To answer this questions, we grouped all papers according to authors' affiliation region, and calculated their T/A dependency of T/A.
We only consider papers with all authors in the same region since it is not clear to attribute dependency to regions with international collaborated papers.
However, we found that single region papers are very similar to international collaborated papers in terms of dependency evolution.
Therefore, the result here will not change much after international collaborated papers are included.
Please see SI for more details.
The results of topmost six regions are shown in \autoref{fig: region_dependency}.
Unlike \autoref{fig: region_trend}, all regions in \autoref{fig: region_dependency} show roughly the same behavior, even for Mainland China and The United States.
This suggests that the trend we found in \autoref{fig: time_dependency} is a universal phenomenon for graphene research and is insensitive to geographic factor.

\section{Conclusion}\label{sec:conclusion}

Discipline-level analysis give many insights of scientific enterprise, also offer important reference for practical purpose, like career decision, hiring decision, funding policy and so on.
Some important works have been done with SciSci paradigm, for example the evolution of sustainability science, the structure and evolution of physics, and the development of artificial intelligence \citep{bettencourt_evolution_2011,sinatra_century_2015,frank_evolution_2019}. 

In this study, we complemented previous studies by investigating evolution of graphene research in terms of its main components: theoretical branch and applied branch.
Using the co-clustering method we divided the graphene publication collection into two groups.
By extracting each group's keywords, we confirmed that one group is about theoretical research and the other does applied research.
Overall, 37.4\% of papers belong to T and remaining 62.6\% are members of A.
However, these ratios are not constants over time: the proportion of T increased to around 90\% in 2007 and gradually decreased afterward, while the proportion of A did just opposite.
It suggests that applied research grown faster than theoretical research and the attention of graphene research community shift gradually from theory to application.
By analyzing authors' affiliations, we computed region credit for every paper and all regions total contribution.
The distribution of contribution is very different in T and A: many regions made significant contribution in T while Mainland China is dominant in A.
And the evolution curves also show significant difference among regions.
Using the dependency factor we invented, the reliance between theoretical research and applied research is quantified.
We found that such dependency is asymmetric: theoretical research is extremely influenced by itself while applied research benefits from T and A in a more balanced way.
Such dependency relation is very stable for theoretical research while changed significantly for applied research.
And we found this phenomenon is insensitive to geographic factor, which suggests it is a universal process.

Although many interest findings were observed, several important questions remain to be answer.
For instance, graphene papers were classified either as theoretical research or applied research in this study.
However, this dichotomy may fail for some papers since collaboration between theorists and experimentists become common nowadays and it is inaccurate to put those paper either in T or A.
In other words, an overlapping-clusters picture of graphene research may be a better description of reality \cite{}.
Future work should take overlapping into account given the importance of collaboration in modern scientific enterprise.
In this study all graphene papers are treated with equal importance.
This choice simplifies our analysis, but introduces deviation from reality: a few papers receive most attention.
It would be beneficial to incorporate this fact into our framework and new results may give us a better picture of graphene research.

\section{Acknowledgements}
This research is supported by the Singapore Ministry of Education Academic Research Fund, under the grant number MOE2017-T2-2-075.

\section{Author Contributions}
Conceived and designed the analysis: Wenyuan Liu.

Collected the data: Ai Linh Nguyen.

Performed the analysis: Ai Linh Nguyen.

Wrote the paper: Ai Linh Nguyen, Wenyuan Liu and Siew Ann Cheong

\section{Appendix A. Region code}
CN: Mainland China.
US: The United States.
JP: Japan.
DE: Germany.
KR: South Korea.
IR: Iran.
IN: India.
RU: Russia.
GB: The United Kingdom.
FR: France.
ES: Spain.
IT: Italy.
SG: Singapore.
TW: Taiwan.

\bibliography{citations}

\end{document}